\documentclass[a4paper,11pt]{article}
\usepackage{pos}

\title{Status and Perspectives on Axion Searches}
\ShortTitle{Axion Searches}

\author*[a,b]{Maurizio Giannotti}

\affiliation[a]{Centro de Astropartículas y Física de Altas Energías, University of Zaragoza, Zaragoza, 50009, Aragón, Spain}
\affiliation[b]{Physical Sciences, Barry University, 11300 NE 2nd Ave., Miami Shores, FL 33161, USA}

\emailAdd{mgiannotti@unizar.es}

\abstract{
The search for axions and axion-like particles (ALPs) remains a major endeavor in modern physics investigation. 
Axions play essential roles in the quest to understand dark matter, the strong CP problem, and various astrophysical phenomena. 
This paper provides a very brief overview of the current status of experimental efforts, highlighting significant advancements, ongoing projects, and future opportunities. Particular attention is given to cavity haloscopes, helioscopes, and laboratory-based light-shining-through-wall experiments, as well as astrophysical probes. 
Some future perspectives are also discussed.}

\FullConference{%
  2nd General Meeting of COST Action COSMIC WISPers\\
  3-6 September 2024\\
  Istanbul, Turkey
}

\begin{document}
\maketitle

\section{Introduction}

Axions and, more generally, Axion-Like Particles (ALPs) are prominent in modern theoretical physics as they emerge in many extensions of the Standard Model (SM)~\cite{Ringwald:2014vqa}. 
They are generally defined as pseudoscalar particles described by the low-energy Lagrangian:
\begin{eqnarray}
    {\mathcal L}_{a}=\frac12 (\partial_\mu a)^2 
    -m_a^2 a^2  - \sum_{f=e,p,n} g_{af} a \bar{\psi}_f\gamma_5\psi_f
    -\frac14 g_{a\gamma} \, a\,F_{\mu\nu} \tilde{F}^{\mu\nu}\,,
    \label{eq:L_a}
\end{eqnarray}
where $a$ is the ALP field, $m_a$ its mass, $g_{af}$ and $g_{a\gamma}$ parameterize its couplings to fermions ($\psi_f$) and photons, and $F$ and $\tilde{F}$ represent the electromagnetic field tensor and its dual. 
The most prominent example of ALP is the QCD axion, a key prediction of the Peccei-Quinn solution to the strong CP problem~\cite{Weinberg:1977ma,Wilczek:1977pj,Peccei:1977hh} and a favored dark matter (DM) candidate~\cite{Marsh:2015xka,DiLuzio:2020wdo,Adams:2022pbo}.
Beyond QCD axions, ALPs naturally emerge in other SM extensions, such as string theory~\cite{Svrcek:2006yi,Arvanitaki:2009fg,Cicoli:2012sz}, and often exhibit weaker constraints, lacking specific mass-coupling relations. 

The experimental search for axions and ALPs has seen a significant resurgence, driven by innovative techniques capable of probing previously unexplored regions of the parameter space~\cite{Irastorza:2018dyq,Sikivie:2020zpn,Irastorza:2021tdu}. 
Many of these regions are motivated by astrophysical and cosmological considerations, making the prospects for near-term discovery particularly promising~\cite{Giannotti:2017law,Giannotti:2022euq,Carenza:2024ehj}. 
This review provides a concise overview of the recent progress in axion and ALP research, with a focus on the axion-photon coupling, while highlighting key developments and near-future opportunities. 
For technical details, readers may refer to the comprehensive discussion in Ref.~\cite{Irastorza:2018dyq}, and for a broader set of references, to the informative webpage~\cite{AxionLimits}.

\section{Current Experimental Landscape}
Axion searches through the axion-photon coupling are broadly divided into 3 categories, depending on the specific axion source:
\begin{itemize}
\item \textit{Pure laboratory experiments}, such as light shining through a wall (LSW). In this case, axions are produced in a laboratory setting. 
These searches are appealing due to the complete control over the source and minimal assumptions about production mechanisms.
Howeever, their sensitivity is limited due to the highly suppressed signal ($\propto g_{a\gamma}^4$).
\item \textit{Helioscopes}. 
In this case, the source (the Sun) is well understood, requiring minimal assumptions about axion production~\cite{Hoof:2021mld}.
An exception is axion production in the solar magnetic field, which relies on much less well-established assumptions~\cite{Guarini:2020hps,Caputo:2020quz,OHare:2020wum}.
\item \textit{Haloscopes}. 
They search for dark matter axions. Thus, the sensitivity to the axion coupling is necessarily dependent on the specific assumption about the fraction of dark matter made up by axions, a fact that is not known and that cannot be deduced by current theoretical considerations~\cite{Adams:2022pbo}. 
\end{itemize}

\subsection{Laboratory Searches}
\begin{table}[]
\centering
\caption{Most relevant LSW experiments, with some technical considerations. Data from Ref.~\cite{Irastorza:2018dyq}.}
\label{tab:lab-exp}
\begin{tabular}{|c|c|c|c|c|}
\hline
\textbf{Experiment} & $B(\mathrm{T})$ & $L(\mathrm{m})$ & $g_{a \gamma}\left[\mathrm{GeV}^{-1}\right]$ & \textbf{Notes} \\ \hline
ALPS-I~\cite{Ehret:2010mh}   & 5  & 4.3  & $5 \times 10^{-8}$      & completed      \\ \hline
CROWS~\cite{Betz:2013dza}   & 3  & 0.15 & $9.9 \times 10^{-8}$ & uses microwaves \\ \hline
OSQAR~\cite{OSQAR:2015qdv}    & 9  & 14.3 & $3.5 \times 10^{-8}$    &    dipole magnet from LHC     \\ \hline
ALPS-II~\cite{Ortiz:2020tgs}  & 5  & 100  & $2 \times 10^{-11}$    &  dipole magnet from HERA \\ \hline
JURA~\cite{Lindner2017} & 13 & 426  & $10^{-12}$              & concept        \\ \hline
\end{tabular}
\end{table}
The main protagonists are the LSW experiments, in which axions are produced from a coherent source of photons (Light) interacting with a magnetic field, and detected as photons (Shining) after being reconverted after passing through a region opaque to photons (Wall). 
The forefront LSW experiment is  ALPS II~\cite{Spector:2016vwo}, in DESY, which is expected to probe the axion-photon coupling down to $g_{a\gamma}\approx 2\times 10^2\,{\rm GeV^{-1}}$, for masses below 0.1 meV or so (see Tab.~\ref{tab:lab-exp}).
The first run finished on May 2024 and a related publication is expected for early 2025~\cite{LindnerColloquium}. 
The full optics is expected to be completed in 2025~\cite{LindnerColloquium}. 
Ongoing discussions within the Physics Beyond Colliders (PBC) working group at CERN are considering the possibility of a larger scale LSW experiment, JURA, which would significantly increase the magnetic field length while bringing its intensity up to 13 T~\cite{Lindner2017}. 

Polarization experiments operate on a concept similar to LSW, but focusing on detecting anomalies in light polarization caused by axions or other WISPs. 
The prototype experiment in this category is PVLAS~\cite{DellaValle:2015xxa,Ejlli:2020yhk}, while an improved version, \texttt{VMB@CERN}, is currently under discussion within the PBC~\cite{Kunc:2024eld}. 
However, the feasibility of this upgraded experiment remains uncertain.

\subsection{Solar axion detection}
Currently, the optimal way to detect solar axions is the so called Sikivie Helioscope~\cite{Sikivie:1983ip}, which relies on the conversion of solar axions into photons in a terrestrial magnetic field. 
An advantage of this method with respect to pure laboratory experiments, is the possibility to measure the axion flux produced through different mechanisms, even those not induced by the axion-photon coupling (see, e.g.,~\cite{DiLuzio:2021qct}), as well as the flux of other WISPs~\cite{OShea:2024jjw,OShea:2023gqn}.

The currently most advanced example is the CERN Axion Solar Telescope (CAST), which recently published updated analysis results~\cite{CAST:2024eil}, pushing the limits on the axion-photon coupling below any other experimental results, and probing a wide mass region so far accessible only to astrophysics~\cite{Ayala:2014pea,Dolan:2022kul,Carenza:2024ehj}.
CAST successor, BabyIAXO~\cite{IAXO:2020wwp,IAXO:2024wss}, a prototype of the future full International Axion Observatory (IAXO)~\cite{IAXO:2019mpb}, is planned for construction at DESY and expected to begin operations soon, probing significant new regions of the axion parameter space (Cf. left panel of Fig.\ref{fig:experiments}).

There are other possibilities to look for solar axions, including using large underground detectors, which, however, have limited chances to compete with the very strong astrophysical bounds~\cite{Straniero:2020iyi,Capozzi:2020cbu} in the near future (see, e.g., Ref.~\cite{Carenza:2024ehj} for an overview and for references).
Another compelling option is to make use of the solar magnetic field rather than a terrestrial one, and to detect the produced photons flux with a space-borne x-ray detector. 
An explicit analysis, using the NuSTAR detector, was recently presented in Ref.~\cite{Ruz:2024gkl}. 
The advantage of this method relies on the significantly longer magnetic field, which pushes the sensitivity to couplings considerably below the CAST reach, though for a smaller mass range.

\subsection{Haloscopes}

Cavity haloscopes search for dark matter axions using various advanced technologies.
%
The most established is the resonant cavity~\cite{Sikivie:1983ip}, 
which employs the conversion of DM axions in a strong magnetic field that permeates a resonant cavity (see Ref.~\cite{Sikivie:2020zpn} for a recent review).
\begin{figure}
    \centering
    \includegraphics[width=0.48\linewidth]{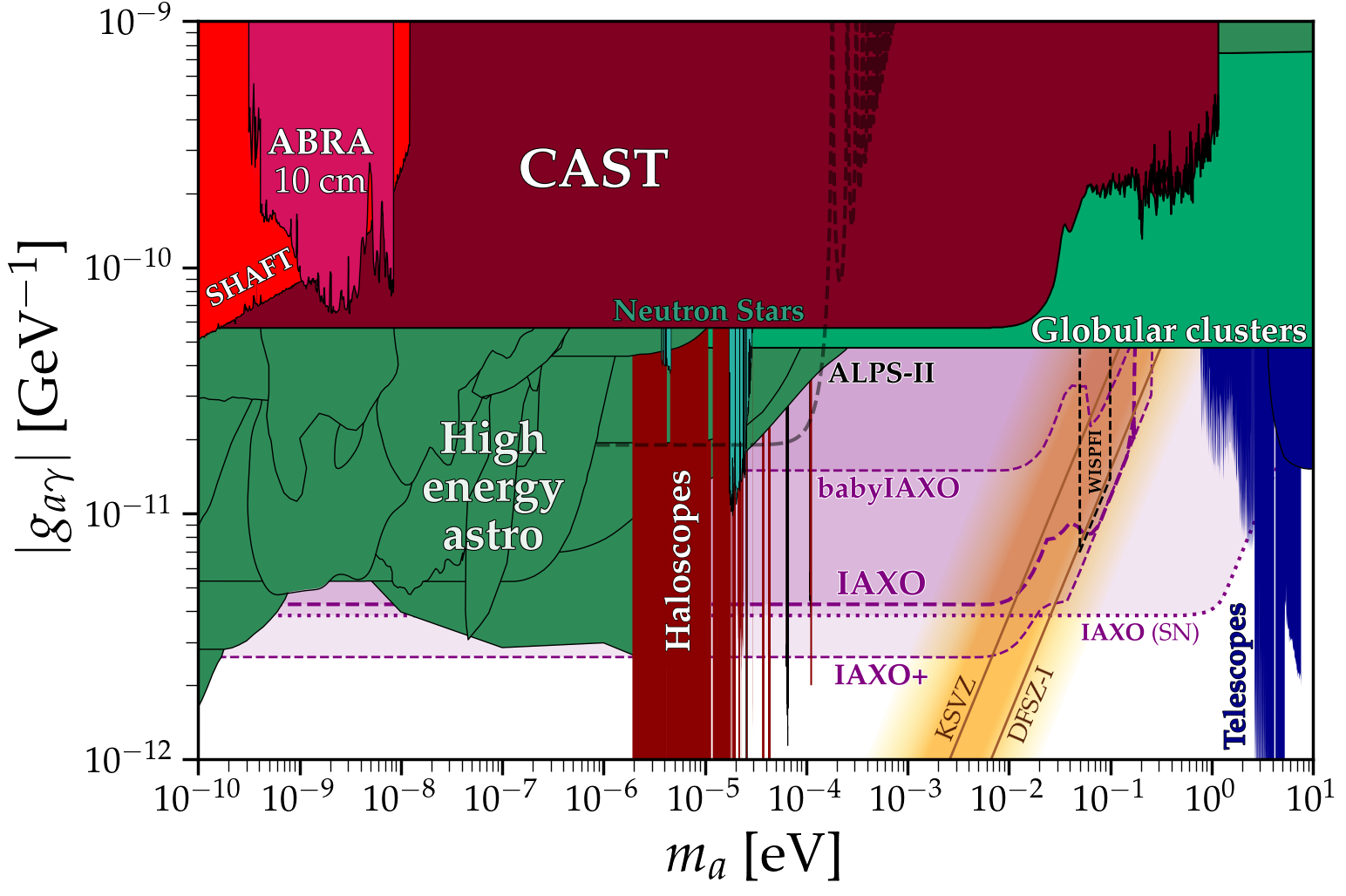}
    \includegraphics[width=0.48\linewidth]{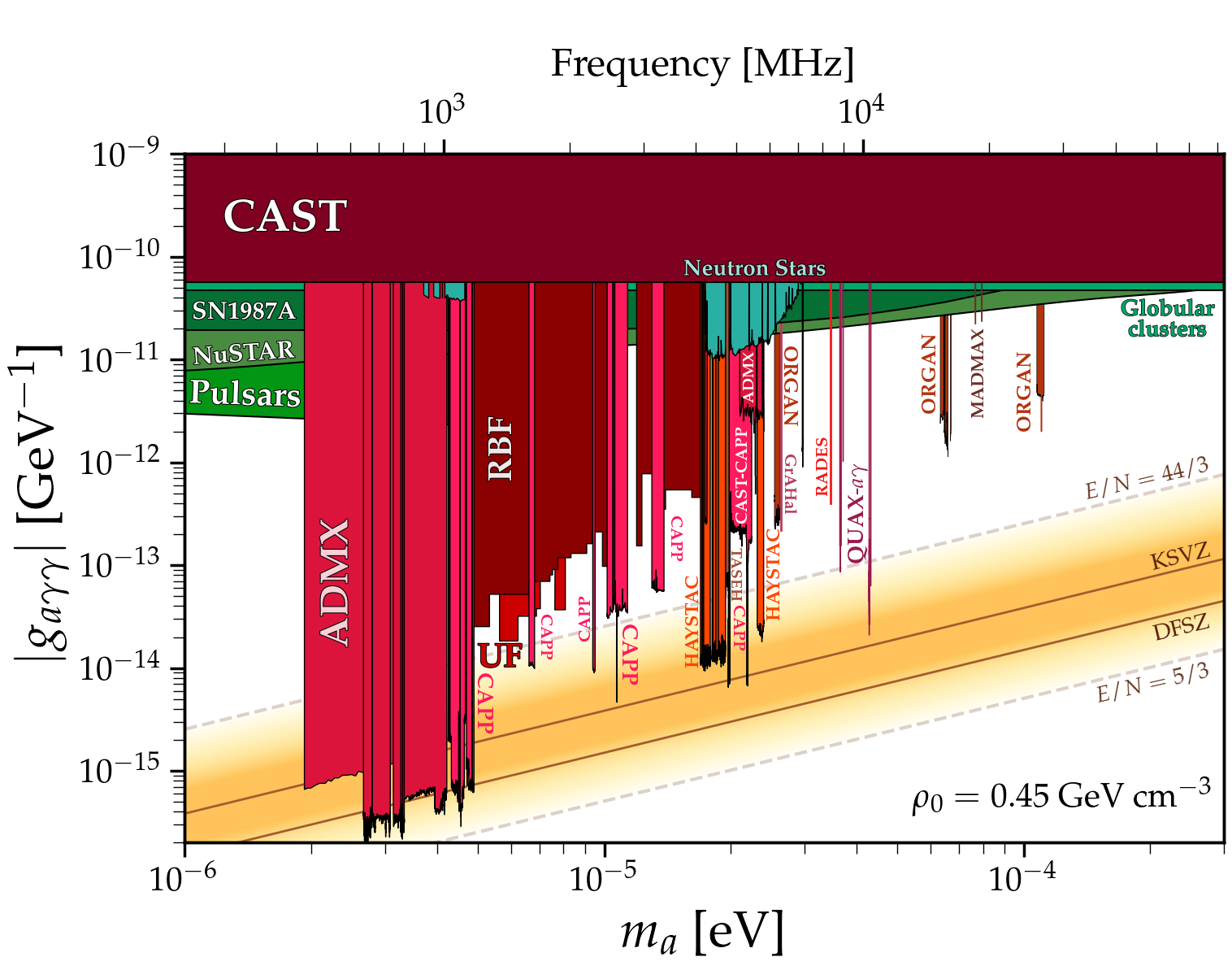}
    \caption{\textit{Left:} Running and proposed axion heliscope experiments;
    \textit{Right:} Currently running cavity experiments, with the reported limits. 
    Notice that the limits on $g_{a\gamma}$ are valid under the assumption that axions constitute the totality of the dark matter in the universe. 
    Details about each experiment, including proper references, can be found in~\cite{AxionLimits}, from which both figures are extracted.}
    \label{fig:experiments}
\end{figure}
Cavities are extremely sensitive instruments (see right panel of Fig.~\ref{fig:experiments}) but present a series of difficulties.
A significant drawback is that, due to the resonance condition, cavities can probe only narrow mass ranges during each scan. 
Furthermore, scanning the parameter space becomes increasingly demanding, as the required time scales quadratically with the desired signal-to-noise ratio ($S/N$), $\Delta t \propto (S/N)^2$.
Furthermore, extending the mass range significantly beyond the region around a few $\mu {\rm eV}$ turns out to be quite challenging. 
The need to overcome these issues has led to a proliferation of very interesting ideas in the latest years. 

ADMX ($B \sim 8 \mathrm{~T}, V \sim 200$ liters), the most mature axion haloscope, has reached the standard QCD band in the mass range $m_a\in [2.7 - 4.2 ]\mu{\rm eV}$~\cite{ADMX:2018gho,ADMX:2019uok,ADMX:2021nhd} and is currently planning to push its sensitivity to higher masses through multicavity designs.
There are plans for the construction of ADMX - EFR (Extended Frequency Reach), 
which might operate at a mass about 1 order of magnitude larger that the one currently under consideration~\cite{RybkaGGI2023}. 
Other cavity experiments are currently probing substantial different masses. 
These include various experiments at the Center for Axion and Precision Physics (CAPP), as shown in Fig.~\ref{fig:experiments}. 
One of the CAPP experiments has recently probed the QCD axion band down to DFSZ sensitivity over a small range of frequencies near $4.55 \mu \mathrm{eV}$~\cite{Yi:2022fmn}, just above the ADMX reach, while HAYSTACK~\cite{Brubaker:2016ktl} is exploring masses about one order of magnitude larger than ADMX.
Other experiments include QUAX--$a\gamma$, probing the mass around $m_a=43~\mu$eV~\cite{Alesini:2020vny}, 
RADES, which has provided a limit at $m_a=34.67 \mu$eV~\cite{CAST:2020rlf},
and ORGAN~\cite{McAllister:2017lkb}, which has a considerably higher mass target $\sim 60-210\mu$eV. 
Considerably different (non-cavity) haloscope concepts are currently operating at lower masses.	
In particular, BASE~\cite{Devlin:2021fpq}, SHAFT~\cite{Gramolin:2020ict}, and ABRACADABRA~\cite{Ouellet:2018beu}
have already reported results in the mass range $m_a\sim 10^{-11}- 10^{-8}\,$eV.

\section{The Road Ahead}
\label{sec:future}

Experimental searches for axions and ALPs have made significant strides in recent years, with numerous new experiments proposed and expected to begin within this decade. 
The current proliferation of experimental proposals brings substantial optimism for exploring a large portion of the axion parameter space in the near future. The outlook is particularly promising for haloscope searches. Projects like MADMAX~\cite{TheMADMAXWorkingGroup:2016hpc,Brun:2019lyf}, CADEx~\cite{Aja:2022csb}, and ultra-low-temperature cavity experiments at IBS/CAPP (CULTASK)\cite{Semertzidis:2019gkj} are set to probe the axion and ALP mass range from the $\mu$eV up to a few 0.1 meV. 
Tunable axion plasma haloscopes~\cite{Lawson:2019brd}, which leverage the axion-plasmon coupling, complement these efforts in the same mass region. 
Meanwhile, proposals such as TOORAD~\cite{Marsh:2018dlj,Schutte-Engel:2021bqm} and BREAD~\cite{BREAD:2021tpx} aim to reach even higher masses, extending the search into the meV regime.

Significant progress has also been made in investigating lower mass regions. The KLASH proposal~\cite{Alesini:2019nzq}, initially designed to target axion masses in the range $m_a \in [0.3-1]\mu$eV, has evolved into FLASH~\cite{Alesini:2023qed}, which explores slightly higher masses due to changes in the magnet configuration~\cite{Gatti:2021cel}. FLASH is poised to investigate not only the axion parameter space but also other WISPs and gravitational waves~\cite{Gatti:2024mde,Visinelli:2024tyw}. Additionally, ongoing experiments like ABRACADABRA are expected to achieve high sensitivity, potentially reaching the QCD axion band in the sub-$\mu$eV range~\cite{Kahn:2016aff}.
Laboratory searches are also progressing. ALPS II~\cite{Bahre:2013ywa} and upcoming axion helioscopes, including BabyIAXO~\cite{IAXO:2020wwp,Dafni:2021mqa,IAXO:2024wss} and IAXO~\cite{IAXO:2019mpb}, aim to explore regions of parameter space below current astrophysical bounds.

What era could an axion discovery usher in?
The discovery of an ALP would not only unveil a new particle and physics at a novel energy scale but also have profound phenomenological implications. 
Determining an axion mass could facilitate its detection through other experimental approaches, particularly using highly sensitive cavity searches, if the mass falls within an accessible range. If a cavity experiment were to confirm a signal previously identified in a laboratory or solar search, it could precisely determine the ALP dark matter fraction—a groundbreaking result.\footnote{Knowing the mass could also enhance LSW experiment sensitivities beyond their nominal capabilities~\cite{Hoof:2024gfk}.}

Moreover, if the axion couplings turn out to be strong enough to be detectable by IAXO—or potentially even BabyIAXO—the particle’s interactions could have measurable impacts on stellar evolution~\cite{Giannotti:2015kwo,Giannotti:2017hny,DiLuzio:2021ysg}.
In this case, axions could also serve as astrophysical messengers, offering unprecedented insights into stellar interiors through methods inaccessible to traditional telescopes~\cite{Carenza:2024ehj}. 
These includes exploring solar properties like temperature~\cite{Hoof:2023jol}, metallicity~\cite{Jaeckel:2019xpa} and magnetic field profiles~\cite{OHare:2020wum,Hoof:2017ibo}, facilitated by anticipated technological advancements.

Beyond the Sun, axions offer remarkable opportunities for studying other stars. 
They could serve as invaluable tools for examining supergiants, providing insights into their evolutionary stages, core temperatures, and other parameters inaccessible through conventional methods~\cite{Xiao:2022rxk,Xiao:2020pra}. 
Similarly, axions might help explore supernova physics, potentially permitting a deeper understanding of the equation of state of nuclear matter~\cite{Arias-Aragon:2024gdz,Carenza:2018jjc,Lella:2022uwi,Lella:2023bfb,Calore:2023srn,Carenza:2023wsm,Lella:2024hfk}.

\section{Conclusions}
Axion research has made remarkable progress, with significant advancements across experimental, theoretical, and observational fronts. The upcoming decade holds immense potential for breakthroughs, making axion searches one of the most exciting fields in modern physics.

The diversity of experimental approaches—ranging from haloscopes and helioscopes to laboratory based techniques—ensures a comprehensive exploration of the parameter space. These methods are complemented by astrophysical probes that provide critical constraints and opportunities to uncover axion-related phenomena in stellar environments. The synergy between these techniques promises not only to refine our understanding of axion properties but also to push the boundaries of sensitivity to previously unexplored regions.

The next generation of experiments are poised to revolutionize the field. 
By leveraging cutting-edge technology and innovative experimental designs, these efforts aim to probe deeper into the axion parameter space and potentially reveal the particle’s role in the dark matter puzzle, the strong CP problem, and stellar physics.

Beyond the technical milestones, an axion discovery would mark the dawn of a new era in physics. Such a breakthrough would unveil physics beyond the Standard Model, opening the door to understanding fundamental questions about the universe. Moreover, it would provide a unique tool to investigate astrophysical phenomena, offering insights into stellar interiors, supernova dynamics, and stellar magnetic field structures that are otherwise inaccessible.

The progress achieved so far underscores the transformative potential of axion research. As we move forward, the collaborative efforts of the scientific community, supported by advancements in technology and theoretical understanding, bring us closer to unraveling one of the most profound mysteries in modern science.

\section*{Acknowledgments}
I express my profound gratitude to the local organizing committee for their excellent organization of the 2nd General Meeting of the COST Action CA21106, at the Istinye University, in Istanbul. 
I acknowledge support from the Spanish Agencia Estatal de Investigación through grant PID2019-108122GB-C31, funded by MCIN/AEI/10.13039/501100011033, and the “European Union NextGenerationEU/PRTR” (Planes complementarios, Programa de Astrofísica y Física de Altas Energías). Additional support comes from grant PGC2022-126078NB-C21, “Aún más allá de los modelos estándar,” funded by MCIN/AEI/10.13039/501100011033 and “ERDF A way of making Europe,” as well as the European Union’s Horizon 2020 research and innovation programme under the European Research Council (ERC) grant agreement ERC-2017-AdG788781 (IAXO+). This work is also based on contributions from COST Action COSMIC WISPers CA21106, supported by COST (European Cooperation in Science and Technology).

\end{document}